\documentclass{iopart}
\usepackage{iopams}
\usepackage[dvips]{graphicx}
\begin{document}
\title{ Exotica in rotating compact stars}
\author{Debarati Chatterjee, Sarmistha Banik, 
and Debades Bandyopadhyay}
\address{Saha Institute of Nuclear Physics, 1/AF Bidhannagar, 
Kolkata 700 064, India}

\begin{abstract}
The determination of mass and radius of a single neutron star 
EXO 0748-676 has been reported recently. Also, the estimate of radius from the 
measurement of moment of inertia of pulsar A in double
pulsar system PSR J0737-3039 would be possible in near future. Here we construct
models of static and uniformly rotating neutron stars involving exotic matter 
and compare our theoretical calculations with the recent findings from 
observations to probe dense matter in neutron stars.

\end{abstract} 

\section{Introduction}
Neutron stars are ideal astrophysical laboratories to study properties
of cold and dense matter \cite{Glenb,Web}.
The investigation of dense and cold matter in neutron star core has entered 
into a new phase. Observatories such as Chandra X-ray observatory, 
X-ray Multi Mirror (XMM)-Newton, Rossi X-ray Timing Explorer (RXTE) are
pouring in many new and interesting data on compact stars. It is now possible
to estimate mass and radius of a single neutron star from observations.
These informations may shed light on the composition and equation of state 
(EoS) of matter at low temperature and high baryon chemical potential.

A lot of interest has been generated in thermonuclear x-ray bursts 
from the low mass x-ray binary EXO 0748-676. 
Already Cottam et al. \cite{Cot} estimated
the gravitational redshift $z=$ 0.35 of three absorption lines in x-ray bursts
from EXO 0748-676 with XMM-Newton. They also reported the narrow widths of
those lines. Mass to radius ratio is immediately known from the measurement of 
$z = (1 - 2GM/{c^2} R)^{-1/2} - 1$, where M and R are neutron star mass and 
radius. 
On the other hand, the widths of absorption lines provide information about
the surface rotational velocity ($v_{rot}$) which is related to 
radius and spin frequency $v_{rot} \propto \nu_{spin} R$. 
Villarreal and Strohmayer \cite{Vil} reported the neutron star spin frequency 
($\nu_{spin}$) of 45 Hz in the x-ray burst oscillations of EXO 0748-676 with
RXTE. These observations lead to a radius of 9.5-15 km and mass 
1.5-2.3M$_{\odot}$ for EXO 0748-676 \cite{Vil}. For the first time, it has 
been possible to determine mass and radius of a single neutron star. 

Recently discovered double pulsar system PSR J0737-3039 \cite{Ly} provides 
another
opportunity for the determination of mass and radius of same neutron star.
Two pulsars in this double pulsar system are denoted by pulsar A and pulsar B.
Pulsar A has a spin period of $\sim$ 23 ms and mass 1.34M$_{\odot}$ whereas 
those of
pulsar B are 2.8 s and 1.25M$_{\odot}$. Double neutron star binaries provide 
unique laboratories for testing relativistic gravity. Soon it will be possible 
to measure spin-orbit coupling of pulsar A from pulsar timing data 
\cite{Mor,Lat}. The moment of inertia for pulsar A could be determined from 
the measurement of spin-orbit 
coupling. Moment of inertia is dimensionally proportional to mass times
radius squared. Already, the mass of pulsar A is known. Therefore, the 
measurement of moment of inertia will provide an estimate of radius of the 
compact object and put important constrain on the properties of neutron star 
matter.

Now the observed mass-radius (M-R) relationship can be directly compared with
that of theoretical calculation. Consequently, the composition and EoS of dense 
matter in neutron stars may be probed. Here we discuss exotic 
forms of matter and their influence on various properties of static and 
rotating compact stars. In section 2, we describe the models of static and 
uniformly rotating compact stars and EoS adopted for this calculation. Results 
are discussed in section 3. And section 4 is devoted to summary and conclusions.

\section{Model}
Models of static and rotating compact stars are
constructed by solving Einstein's field equations.
The equilibrium structures
of non-rotating stars are obtained from Tolman-Oppenheimer-Volkoff (TOV)
equations.
For rotating neutron stars, we consider stationary and axisymmetric equilibrium
configurations. In this case, the metric is taken in the form \cite{Cook,Ster}
\begin{equation}
ds^2 = - e^{\gamma + \rho} dt^2 + e^{2\alpha} (dr^2 + r^2 d{\theta}^2)
       + e^{\gamma - \rho} r^2 sin^2{\theta}(d\phi - \omega dt)^2 ~,
\end{equation}
where $\omega$ is the angular velocity of the local inertial frame. The metric
functions in the line element depend on the radial coordinate $r$ and the polar
angle $\theta$.

We model the neutron star matter as a perfect fluid and the stress-energy 
tensor is given by
\begin{equation}
T^{\mu \nu} = (\varepsilon + P)u^{\mu}u^{\nu} - P\,g^{\mu \nu}
\end{equation}
where $\varepsilon$ and $P$ are the total energy density and pressure of the
matter respectively and $u^a$ is the fluid's four velocity. The moment of 
inertia of the star is defined by $I = J/{\Omega}$; angular momentum ($J$)
is,
\begin{equation}
J = \int {d^3x{\sqrt(-g)}(\epsilon + P) u^t (g_{\phi\phi} u^{\phi} 
+ g_{\phi t} u^t)}
\end{equation}
where, $u^{\phi}$ and $u^{t}$ are components of fluid four velocity and 
$\Omega$ is angular velocity of the star.
For uniformly rotating neutron stars, we adopt a numerical code based on 
Komatsu-Eriguchi-Hachisu \cite{Ster,kom} method. 

Composition and structure of neutron stars depend on the nature of strong 
interaction.
Matter density could exceed by a few times normal nuclear matter density in a 
neutron star interior. Consequently, the baryon and lepton chemical potential
increase rapidly with baryon density. This might lead to the formation of exotic
matter with a large strangeness fraction such as hyperon matter, Bose-Einstein
condensate of $K^-$ mesons and quark matter. Here we discuss equations of state
with various compositions of dense matter in neutron star cores. 

We construct hadronic EoS using relativistic field theoretical models 
\cite{Bani1,Bani2}. Here baryon-baryon interaction is mediated by exchange
of mesons. The hadronic phase is composed of all members of the baryon octet,
electrons and muons. The hadronic phase is to satisfy 
beta-equilibrium and charge neutrality conditions. In compact stars interior, 
the generalised $\beta$-decay processes may be written in the form
$B_1 \longrightarrow B_2 + l+ \bar \nu_l$ and 
$B_2 +l \longrightarrow B_1 +\nu_l$ where $B_1$ and $B_2$ are baryons and 
l represents leptons. The generic equation for chemical equilibrium condition is
\begin{equation}
\mu_i = b_i \mu_n - q_i \mu_e ~,
\end{equation}
where $\mu_n$, $\mu_e$ and $\mu_i$ are respectively
the chemical potentials of neutrons, electrons and i-th baryon and $b_i$ and 
$q_i$ are baryon number and electric charge of i-th baryon respectively. The  
charge neutrality in hadronic phase is given by 
$Q^h = \sum_B q_B n^h_B -n_e -n_\mu = 0$, where
$n_B^h$ is the number density of baryon B in pure 
hadronic phase and $n_e$ and $ n_\mu$ are 
number densities of electrons and muons respectively.
 
Next we consider Bose-Einstein condensate of $K^-$ mesons. 
Experimental results from heavy ion collisions suggest that the in-medium 
kaon-nucleon interaction is repulsive whereas it is attractive for 
antikaon-nucleon \cite{Pal,Li}. Also, informations about antikaon-nucleon 
interaction in medium may be obtained studying the $K^-$-atomic data. The 
analysis of $K^-$-atomic data using a phenomenological density dependent 
potential showed that the real part of antikaon optical potential could be as 
large as $-180\pm 20$ MeV \cite{Fri}.  

It was first demonstrated
by Kaplan and Nelson within a chiral $SU(3)_L \times SU(3)_R$ model that $K^-$ 
meson may undergo Bose-Einstein condensation in dense matter formed in heavy ion
collisions \cite{Kap}. The effective mass of antikaons decreases with 
increasing 
density because of the strongly attractive $K^-$-baryon interaction in dense
matter. Consequently, the in-medium energy of $K^-$ mesons in the zero-momentum
state also decreases with density. The $s$-wave $K^-$ condensation sets in 
when the energy of $K^-$ mesons equals to its chemical potential. 

We consider $K^-$ condensation as a first order phase transition. Here we adopt
a relativistic field theoretical model to describe (anti)kaon-baryon 
interaction \cite{Bani1,Bani2,Gle99}. In this case, (anti)kaon-baryon 
interaction is treated in the same footing as baryon-baryon interaction.
The pure $K^-$ condensed phase is composed of members of the baryon octet 
embedded in the 
condensate, electrons and muons and maintains charge neutrality and 
beta-equilibrium conditions. The charge neutrality condition is  
$Q^{\bar K}=\sum_B q_B n_B^{\bar K} -n_{K^-} - n_e - n_\mu =0$,
where $n_B^{\bar K}$ is the number density of baryon B in pure 
antikaon condensed phase and $n_{K^-}$ is the number density of antikaons.

Strangeness changing processes such as,
$n \rightarrow p + K^-$ and $e^- \rightarrow K^- + \nu_e$,
in a neutron star interior are responsible for the onset of $K^-$ condensation.
The requirement of chemical equilibrium yields
\begin{eqnarray}
\mu_n - \mu_p &=& \mu_{K^-} = \mu_e ~, \\
\end{eqnarray}
where $\mu_{K^-}$ is the chemical
potentials of $K^-$ mesons. The above condition determines the
onset of $K^-$ condensation. 

The deconfinement phase transition from hadronic to quark matter is a 
possibility in a neutron star interior.
In this calculation, we also consider a first order phase transition from 
hadronic to quark phase. The pure quark matter is composed of $u$, $d$ and $s$ 
quarks. The EoS of pure quark matter is described in the MIT bag model 
\cite{Far}.  

As we are considering first order phase transitions, the mixed phase of 
two pure phases is governed
by Gibbs phase rules and global conservation laws \cite{Glenb}. The Gibbs phase 
rules read, $P^I = P^{II}$ and $\mu_B^I = \mu_B^{II}$ where $P^I$, $P^{II}$ and
$\mu_B^I$, $\mu_B^{II}$ are pressure and chemical potentials of baryon B in 
phase I and II respectively. The conditions for global charge neutrality and 
baryon number conservation are respectively $(1-\chi) Q^I + \chi Q^{II}=0$ and
$n_B = (1-\chi) n_B^I + \chi n_B^{II}$ where $\chi$ is the volume fraction
in phase II. 
The total energy density in the mixed phase is given 
by $\epsilon = (1-\chi) \epsilon^I + \chi \epsilon^{II}$.
  
\begin{figure}[t]
\begin{center}
\includegraphics[height=8cm]{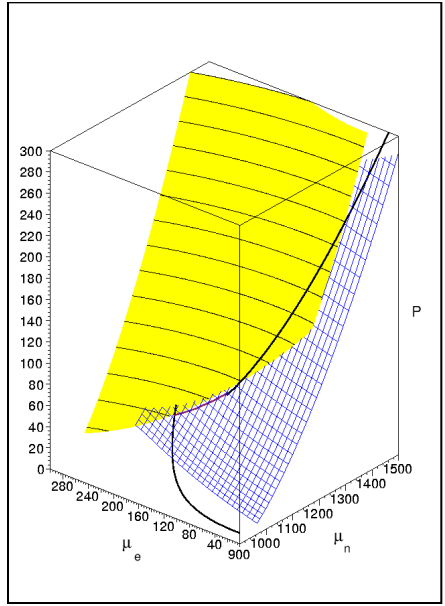} 
\caption{Pressure $P$ (MeV fm$^{-3})$ as a function of $\mu_n$ (MeV) and 
$\mu_e$ (MeV) for
pure hadronic phase (wired surface) and $K^-$ condensed phase (shaded surface).
The solid line is the charge neutrality line in pure hadronic phase, the mixed
phase and pure $K^-$ condensed phase.}
\end{center}
\end{figure}

\section{Results \& Discussions}
We adopt GM1 parameter set \cite{Bani1,Bani2,Gle2} for this calculation where 
nucleon-meson coupling constants are determined from the nuclear matter 
saturation properties. The vector meson coupling constants for 
(anti)kaons and hyperons are determined from the quark model \cite{Sch96} 
whereas the scalar meson coupling constants for hyperons and antikaons are 
obtained from the potential depths of  hyperons and antikaons in normal nuclear 
matter \cite{Bani2,Sch96}. As the phenomenological fit to the $K^-$ atomic 
data yielded a very strong real part of antikaon potential 
$U_{K^-}$ at normal nuclear matter density, we 
perform this calculation with an antikaon optical potential of -160 MeV at 
normal nuclear matter density ($n_0=0.153 fm^{-3}$). The coupling constants 
for strange mesons with hyperons and (anti)kaons are taken from
Ref.\cite{Bani2,Sch96}. Also, we consider a bag constant $B^{1/4}$ = 200 MeV
and strange quark mass $m_s$ = 150 MeV for quark matter EoS.

We compute equations of state for four different compositions of neutron star 
matter. The EoS with
$n$, $p$, $e$ and $\mu$ is denoted by "np". With further inclusion of 
hyperons, it becomes "npH". The EoS undergoing a first order phase transition
from nuclear to quark matter is represented by "npQ". In the last case,
the EoS involves first order phase transitions from nuclear to $K^-$
condensed matter and then from $K^-$ condensed matter to quark matter and is
referred to as "npKQ" \cite{Bani4}.

It is evident from Eq. (4) that neutron and electron chemical potentials are
two independent quantities. One can express energy density and pressure of
neutron star matter using $\mu_n$ and $\mu_e$. As an example, we demonstrate
here the construction of EoS involving a first order phase
transition from nuclear to $K^-$ condensed matter. In Figure 1, pressure is 
plotted as a function of neutron and electron chemical potentials for pure 
hadronic and $K^-$ condensed phase. Here shaded surface
denotes the antikaon condensed phase whereas the wired surface represents
the hadronic phase. It is noted that two surfaces intersect along a
common line. This implies that two phases having same pressure along the 
intersection line could co-exist. In this figure, the solid line represents 
the EoS 
of charge neutral and beta-equilibrated matter. It has three parts. Below 
P=42 MeV fm$^{-3}$, the pressure in pure hadronic phase is higher than 
that of the antikaon condensed phase making the hadronic phase physically 
preferred one. Therefore, the solid line below 
pressure 42 MeV fm$^{-3}$ denotes the charge neutrality line in pure hadronic 
phase. This phase is mainly composed of neutrons, protons, electrons and muons. 
Two surfaces first meet at 2.23$n_0$ and P=42 MeV fm$^{-3}$ and 
this is the beginning of the mixed phase. After the onset of the 
phase transition, the mixed phase represented by the solid line follows the 
intersection part of two surfaces. The Gibbs conditions of phase equilibrium 
and global charge neutrality are satisfied along the intersection line of two 
surfaces. We find that the electron chemical potential $\mu_e$ decreases in the 
mixed phase. This is attributed to the fact that with the formation of $K^-$ 
condensate, electrons are replaced by $K^-$ mesons. The phase transition 
ends at 3.51$n_0$ and pressure 70 MeV fm$^{-3}$. The solid line above 
pressure 70 MeV fm$^{-3}$ represents the charge neutrality line in pure 
antikaon condensed phase. 

\begin{figure}[t]
\begin{center}
\includegraphics[height=8cm]{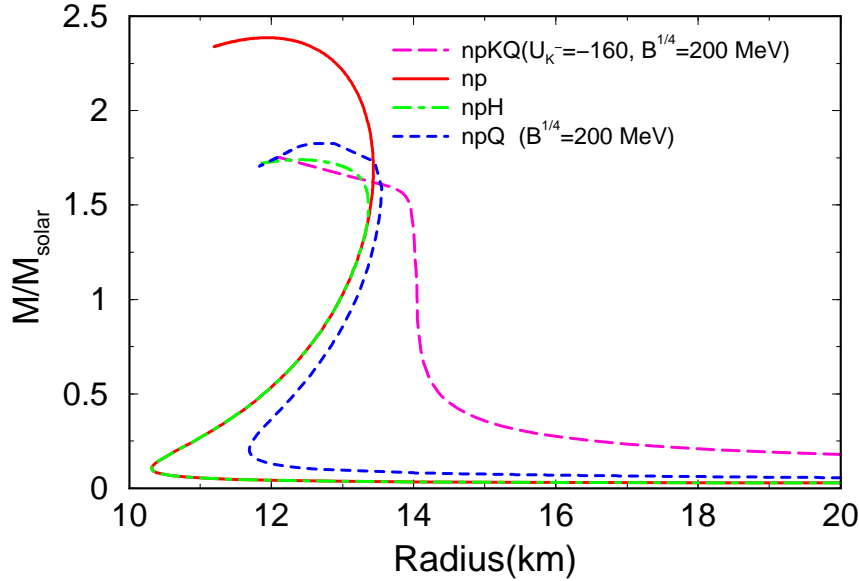} 
\caption{Mass-radius relationship of non-rotating stars calculated with 
different equations of state.}
\end{center}
\end{figure}
In Figure 2, the mass-radius relationship of non-rotating neutron stars is 
displayed for different EoS. For np case, the maximum neutron star mass is 
2.39 M$_{\odot}$
and the corresponding radius is 11.94 km. With the inclusion of softening 
components such as hyperons, Bose-Einstein condensate of $K^-$ mesons or 
quarks, the 
maximum neutron star masses are reduced. For npH case, the maximum mass and
the corresponding radius are 1.74M$_{\odot}$ and 12.41 km respectively. 
Maximum masses and the corresponding radii are 
1.83 M$_{\odot}$ and 12.88 km for npQ case and 1.75M$_{\odot}$ and 12.10 km
for npKQ case, respectively. Among all cases considered here, the maximum
neutron star mass for np case is the largest because the EoS in this case is 
the stiffest. Now we compare our results with the recent
findings from EXO 0748-676. Villarreal and Strohmayer \cite{Vil} have predicted
that the best fit values of mass and radius for EXO 0748-676 are M$\sim$ 1.8 
M$_{\odot}$ and R$\sim$ 11.5 km respectively. It is found that theoretical 
M-R relationships shown in Figure 2 are consistent with the observed values. It 
shows that there might be room for exotic matter in neutron star interior. 

\begin{figure}[t]
\begin{center}
\includegraphics[height=8cm]{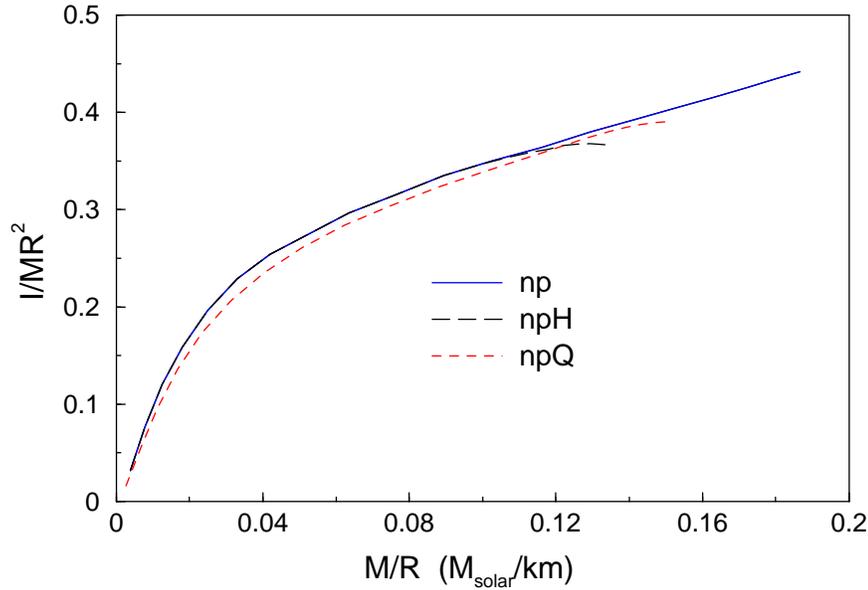} 
\caption{Moment of inertia (I) is plotted with compactness parameter (M/R) for
three different EoS.}
\end{center}
\end{figure}

Now we focus on the estimate of radius from the calculation of moment of 
inertia of pulsar A in the double pulsar binary PSR J0737-3039. In this 
calculation, we adopt some of EoS used for explaining EXO 0748-676 data.
Also, we use the inputs gravitational mass M=1.337M$_{\odot}$ and angular
velocity $\Omega$ = 276.8 s$^{-1}$ corresponding to a spin period 
P=${2\pi}/{\Omega}$=22.7 ms to calculate the moment of inertia of the uniformly 
rotating star. It follows from the calculation that for np, npH and npQ cases 
the central energy density corresponding to pulsar A is $\sim$ 5.4 $\times$ 
10$^{14}$ g cm$^{-3}$. We also note that the moment of inertia and radius of 
pulsar A are I $\sim$ 4.90 $\times$ 10$^{45}$ g cm$^2$ and R $\sim$ 14 km 
for all these cases studied here. 
It exhibits that hyperons or quark matter does not change the moment of inertia
and radius appreciably. 

In Figure 3, we have shown the moment of inertia in dimensionless unit as a
function of compactness parameter (M/R) of the neutron star for different
EoS. Already, it has been noted that these EoS give rise to maximum neutron
star masses
$>$ 1.7 M$_{\odot}$. Therefore, $I/{MR^2}$ versus M/R curves shown in Figure 3
can be approximated by \cite{Lat}  
\begin{equation}
{I \simeq (0.237 \pm 0.008)\, MR^2 \>\Biggl[ 1+ 4.2 \>\frac{M\,km}{M_\odot\,R}
+90 \>\Biggl(\frac{M\,km}{M_\odot\,R}\Biggr)^4 \,\Biggr]}~.
\end{equation}

The values of moment of inertia that we have obtained from the rotating neutron
star model for fixed angular frequency are comparable with the values 
calculated using Eq. (7). Thus from the measurements of moment of inertia and 
mass of pulsar A, the radius could be determined by inversion of Eq. (7).

It has been shown that the moment of inertia of a neutron star depends quite 
strongly on the EoS. 
The predictions for neutron star moment of inertia and radius are quite
different in non-relativistic and relativistic models \cite{Mor}. 
Therefore, a measurement of moment of inertia with 10 percent uncertainty 
could impose new constraints on the EoS and could rule out some classes of
equations of state that currently exist.

\section{Summary and conclusions}
We have computed various properties of static and uniformly rotating compact 
stars using EoS with first order nuclear to $K^-$ condensed matter phase 
transition and also $K^-$ condensed matter to quark matter phase transition. 
Mass and radius of EXO 0748-676 have been measured. These informations put
stringent constraints on EoS. Our results are consistent with the mass and
radius of EXO 0748-676. This, in turn, shows that bizarre matter might exist
in this star. We have also calculated the moment of inertia for pulsar A of
double pulsar binary PSR J0737-3039. It is found that the values of moment of
inertia and radius obtained with relativistic EoS are quite different from 
that of the nonrelativistic model of Akmal and Pandharipande \cite{Mor}. Soon 
the measurement of moment of 
inertia for pulsar A would be possible. This would lead to an estimate of the 
radius of pulsar A. Therefore, it would be possible to rule out some EoS with
the help of the measured value of moment of inertia and mass of pulsar A. 
Lattimer and Prakash showed that there was an empirical correlation between
the radius and matter pressure in the density region n$_0$ to 2n$_0$ 
\cite{Pra}. Therefore, the determination of radius of pulsar A having a mass 
M=1.337M$_{\odot}$ could give an estimate of matter pressure around saturation
density. It is interesting to note that we have informations about mass and 
radius of a single neutron star. More events like this would be available in
future. Therefore, it would be worth investigating the
dense matter EoS from the knowledge of masses and radii of neutron stars and
using a numerical inversion of neutron star structure equation \cite{Lin}.

\noindent{\bf Acknowledgments:}

\noindent Two authors (SB and DB) acknowledge the support for this work by the 
Department of Science and Technology (DST), Government of India and German 
Academic Exchange Service (DAAD), Germany.

\end{document}